\documentclass{aa}
\usepackage{graphicx}
\usepackage{times}
\usepackage{astron}
\usepackage{psfig}

\def\hide#1{}

\begin{document}

\thesaurus{03.13.2; 03.13.6; 03.20.4; 04.19.1}

\title{Multi-color Classification in the Calar Alto Deep Imaging Survey
\thanks{Partly based on observation collected at the European Southern Observatory, Paranal, Chile (ESO Programmes 64.O-0401}}

\author{C. Wolf\inst{1} \and K. Meisenheimer\inst{1} \and H.-J. R\"oser\inst{1} \and S.V.W. Beckwith\inst{1,2} \and F. H. Chaffee, Jr.\inst{3} \and J. Fried\inst{1} \and H. Hippelein\inst{1} \and J.-S. Huang\inst{1,4} \and M. K\"ummel\inst{1} \and B. von Kuhlmann\inst{1} \and C. Maier\inst{1} \and S. Phleps\inst{1} \and H.-W. Rix\inst{1} \and E. Thommes\inst{1} \and D. Thompson\inst{1,5} }

\institute{
 Max-Planck-Institut f\"ur Astronomie, K\"onigstuhl 17,
   D-69117 Heidelberg, Germany
\and
 Space Telescope Science Institute, 3700 San Martin Drive, Baltimore, MD 21218, USA
\and
 W. M. Keck Observatory, 65-1120 Mamalahoa Highway, Kamuela, Hawaii 96743, USA
\and
 Harvard-Smithsonian Center for Astrophysics, 60 Garden Street, Cambridge, MA 02138, USA
\and
 California Institute of Technology, Pasadena, CA 91125, USA}

\date{Received 20 July 2000 / Accepted }
\maketitle


\begin{abstract}
We use a multi-color classification method introduced by Wolf, Meisenheimer \& R\"oser (2000) to reliably identify stars, galaxies and quasars in the up to 16-dimensional color space provided by the filter set of the Calar Alto Deep Imaging Survey (CADIS). The samples of stars, galaxies and quasars obtained this way have been used for dedicated studies which are published in separate papers. 

The classification is good enough to detect quasars rather completely and efficiently without confirmative spectroscopy. The multi-color redshifts are accurate enough for most statistical applications, e.g. evolutionary studies of the galaxy luminosity function. Also, the separation between stars and galaxies reaches deeper than with morphological criteria, so that studies of the stellar population can be extended to fainter levels.

We characterize the dataset presently available on the CADIS 1h-, 9h- and 16h-fields. Using Monte-Carlo simulations we model the classification performance expected for CADIS. We present a summary of the classification results on the CADIS database and discuss unclassified objects. More than 99\% of the whole catalog sample at $R<22$ (more than 95\% at $R<23$) are successfully classified matching the expectations derived from the simulations. A small number of peculiar objects challenging the classification are discussed in detail.

Spectroscopic observations are used to check the reliability of the multi-color classification (6 mistakes among 151 objects with $R<24$). From these, we also determine the accuracy of the multi-color redshifts which are rather good for galaxies ($\sigma_z \approx 0.03$) and useful for quasars. We find that the classification performance derived from the simulations compares well with results from the real survey. Finally, we locate areas for potential improvement of the classification.

%
\end{abstract}

\section{Introduction}

The {\it Calar Alto Deep Imaging Survey} (CADIS) is an extragalactic key project at the Max-Planck Institut f\"ur Astronomie (MPIA), Heidelberg, which is aiming at two types of objectives: CADIS investigates whole samples of different object classes using statistical tools, but it also searches for individual rare and faint objects, which will be studied in detail with coming large telescopes. As a pencil beam survey, it probes seven different fields at galactic latitudes $b \ga 45\degr$ with a total area of $\sim 0.25\degr$.

The final object catalog will arise from two fundamentally different survey techniques:
\begin{itemize} 
\item   a multi-color survey with B, R, J and K$^\prime$ plus 13 medium-band filters from 400\,nm to 1000\,nm, practically resembling low-resolution imaging spectroscopy and giving a complete list of objects with $R \la 23$, 
\item   and an emission-line survey using an imaging Fabry-Perot interferometer to probe emission line galaxies down to a limiting line flux of $\sim 3 \times 10^{-20} Wm^{-2}$.
\end{itemize}

Presently, the data for three fields are reduced and have been used for a number of application studies published already or to be published this year. For many applications, objects of concern are selected by our multi-color classification which uses the many bands to sort the objects into {\it stars}, {\it galaxies} and {\it quasars}.  Also, multi-color redshifts are estimated for the extragalactic objects. This classification scheme was originally developed on the basis of CADIS data, but is meanwhile used in a range of different survey activities. It uses a library of $\sim 65000$ templates and achieves high classification reliability ($>90\%$ correct classification in each class) and a high redshift accuracy of $\sigma_z \approx 0.03$ for galaxies and $\sigma_z \approx 0.1$ for quasars. The methodogical background for the classification was published by Wolf, Meisenheimer \& R\"oser (2000), hereafter paper I. There, the classification method is derived from statistical principles, the libraries are defined, the performance expected with different filter sets is compared and where conclusions are drawn for optimum survey strategies.

The purpose of this paper is to characterize the present data of the CADIS multi-color survey and to discuss the classification performance which was checked by a subsample of objects with spectroscopic identifications. The paper is organized as follows: Section 2 lists CADIS goals for which the classification is relevant and discusses what kind of objects the classifications should be prepared for. Section 3 defines the present CADIS dataset and characterizes its photometry and calibration. Section 4 outlines the classification method and presents statistics of classified and unclassified objects. Section 5 discusses the classification performance using a spectroscopic cross check sample. Finally, Section 6 summarizes the quality of the classified catalogs and evaluates the practical performance of the CADIS multi-color classification. A few identified peculiar objects challenging the classification are discussed in the appendix.

\section{Objects and objectives in CADIS}

Since the object classification in CADIS is based on colors, we like to ensure that the measured colors are accurate and not reddenned by interstellar dust absorption in our Galaxy. Therefore, all CADIS fields are placed in zero-reddenning areas, i.e. local minima of the IRAS 100\,$\mu$ maps with undetected fluxes ($< 2$\,MJy/sterad). A second obvious advantage of this choice is increased depth for extragalactic work (see Table\,\ref{t-fields} for coordinates of the three CADIS fields included in this paper). 

Although we know what objects we are looking for, we have to carefully determine which objects we might be confronted with in order to check whether a color-based classification can deal with it. A CADIS field measures at least $10\arcmin \times 10\arcmin$ in size when observed with MOSCA at the 3.5-m-telescope on Calar Alto. Most observations used CAFOS at the 2.2-m-telescope and cover a larger round field of 14\arcmin diameter and $154\sq\arcmin$ area. On the minimum area of $10\arcmin \times 10\arcmin$ we expect the following objects in a typical CADIS field:
\begin{itemize}

  \item  Galaxies are the most abundant object class in the CADIS fields. Published number counts let us expect some 750 galaxies with $R<23$ per field \cite{Met95}.

  \item  Stars should be the second-most common objects. According to models of the Milky Way we expect 100 to 150 stars with $R<23$ per field depending on its coordinates \cite{BS81}. Most stars should be late-type main sequence stars, and only few should be of other type: We expect two white dwarfs with $R<23$ by assuming their spatial distribution to follow the general thin plus thick disk population of main sequence stars and by using their local luminosity function \cite{BS93}. Furthermore, we expect 0.15 red disk giants with $V>12$ per field area \cite{BS81}. Also, halo giants should be negligible in the interesting CADIS magnitude range of $18 \la R \la 23$. While T dwarfs are not to be expected with $R<23$ in such a small survey area at high galactic latitudes, L dwarfs could be present in low numbers. In fact, CADIS has identified already an L1-dwarf with $I=22.5$ undetected in $R$ \cite{Wolf98}.

  \item  Active galaxies with broad-line emission spectra are supposedly the third-most common object population in the CADIS fields. At $B<23.5$ we expect 12 objects per field, i.e. about eight Seyfert-1 galaxies and four quasars \cite{HS90}, with the most distant objects residing out to $z \la 3$.

  \item  The total space density of Seyfert-2 galaxies is estimated to be four to eight times higher than Seyfert-1 galaxies \cite{Wol90}. We do not have estimates for their abundance in the CADIS fields. They should be visible in the redshift range of normal galaxies, since their SED is dominated by the host, especially at UV restframe wavelengths. Up to now, we are unable to separate them from starburst galaxies with CADIS photometry. In any case, there should be few objects with $R<23$ per field.

  \item  We do not really expect a BL Lac object in our fields, since only a few hundred of them are known on the entire sky, and CADIS will survey in total $\sim$ 1/200,000 of the sky. 
\end{itemize}

\begin{table}
\caption[CADIS field centers]{\label{t-fields}
Positions of the CADIS field centers on the sky ($\pm5\arcsec$)  and number of filters observed, see Tab.\,2 for details:} 
\medskip
\begin{center}
\begin{tabular}
{l r@{$^{\rm h}$}r@{$^{\rm m}$}r@{\fs}l
r@{\degr}r@{\arcmin}r@{\arcsec\ } ccc}
\hline
\noalign{\smallskip}
\multicolumn{1}{c}{FIELD} &
\multicolumn{4}{c}{$\alpha$ (2000)} &
\multicolumn{3}{c}{$\delta$ (2000)} &
$l_{galactic}$ & $b_{galactic}$ & $N_{filter}$ \\ 
\noalign{\smallskip}
\hline 
\noalign{\smallskip}
\, 1\,h                &  1 & 47 & 33 &  3 &  2 & 19 & 55	& 150\degr & -59\degr & 12 \\ 
\, 9\,h                &  9 & 13 & 47 &  5 & 46 & 14 & 20	& 175\degr &  45\degr & 17 \\ 
16\,h                  & 16 & 24 & 32 &  3 & 55 & 44 & 32	&  85\degr &  45\degr & 13 \\ 
\noalign{\smallskip}
\hline 
\end{tabular}
\end{center}
\end{table}

Therefore, we conclude that classifying objects into {\it stars}, {\it galaxies} and {\it quasars} would be sufficient for classifying more than 99\% of the objects in the catalog. In this terminology, we included all broad emission-line AGNs into the term {\it quasar}. Using these classes, the multi-color object catalog of CADIS has been partitioned into subsamples and used as an input to four applications, for which completeness and contamination by wrongly classified objects can make an important difference:
\begin{itemize}

  \item  Deep star counts are used to probe the galactic structure and the stellar luminosity function. Our multi-color classification can separate stars and galaxies even at faint levels where the morphology is difficult to measure. Consequently, compact galaxies are also eliminated as contaminants from the list of stars. CADIS probes the Milky Way on different lines of sight and allows to derive stellar density distributions. This way a clear signature of a thick disk was found in the first two fields analysed \cite{Phl00}.

  \item Galaxies of $I<23$ and $z\la 1$ were used to investigate the evolution of the galaxy luminosity function with redshift. Such an analysis is particularly sensitive to selection effects and K corrections. By not excluding morphologically stellar objects, we include an extra population of compact galaxies, which at $R=21\ldots23$ and $1\arcsec$ seeing make up 20\% of all galaxies. The many wavebands in the CADIS filter set give us complete SED information and provide accurate restframe B-band photometry directly from the data,  as long as the redshift estimate is correct. This way we do not need estimated K corrections at least out to $z \sim 1$. Also, the accuracy of the multi-color redshifts allows to work with the full multi-color sample of $\sim 10000$ galaxies when completed. The first study with 2779 objects  maps out the luminosity evolution which is clearly differential among the galaxy types where starburst galaxies show a steepening of their luminosity function with redshift and an increasing space density \cite{Fri00}.

  \item  Galaxies from our medium-deep K-band images ($K^\prime < 19.5$) were used for a number count analysis. Again, the selection of galaxies works to rather deep limits and even compact galaxies are taken into account. This study establishes the currently largest statistics among medium-deep K-band surveys and settles earlier divergence on K-band counts in the range of $16.5 < K^\prime < 19.5$ \cite{Hua00}.

  \item  Quasars with $R<22$ were used for a first comparison with expected counts. This rare class is most sensitive to contamination, of which the CADIS classification is almost free. Completeness appears to be extremely high, given that the CADIS counts show an excess in faint high-redshift ($z>2$) quasars. Also, a few interesting quasar pairs at separations on the order of $\sim 1\arcmin - 5\arcmin$ were spectroscopically confirmed \cite{Wolf99}.
\end{itemize}

\section{The CADIS data}

\begin{figure*}
\centerline{\psfig{figure=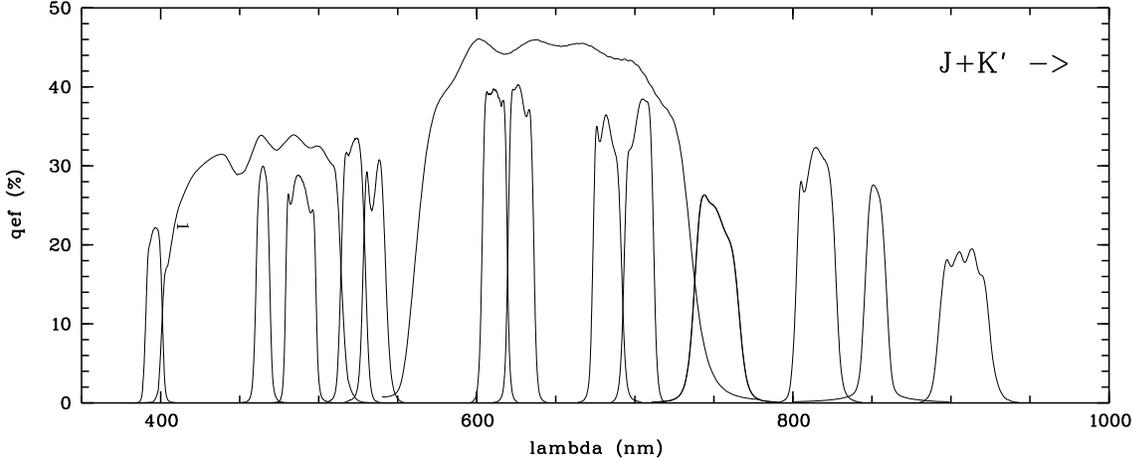,angle=270,clip=t,width=15cm}}
\caption[ ]{The quantum efficiency of all CADIS filters after taking the entire observing system into account. The set contains four broadband filters, CADIS-B, CADIS-R, J and K$^\prime$ (not shown), and 13 mediumband filters. All filters are used for classification and redshift estimation. \label{filterset}}
\end{figure*}

\subsection{Basic data characteristics in CADIS}

The CADIS database contains information on multi-band photometry, morphology and position of each object. The photometry will encompass data from the four broad-band filters CADIS-B, CADIS-R, J and K$^\prime$ as well as from 13 medium-band filters when completed. This photometric database is collected over five years at Calar Alto Observatory in Spain using the focal reducers CAFOS at the 2.2-m-telescope, MOSCA at the 3.5-m-telescope and the prime focus near-infrared camera OMEGAprime at the 3.5-m-telescope \cite{Biz98}. The basic data reduction steps like flatfielding, cosmic correction and coadding dithered frames are done with the MIDAS software package in combination with a dedicated CADIS reduction context based on routines from MPIAPHOT (by Meisenheimer \& R\"oser)

Depending on their spectrum objects are detected in different bands with different signal-to-noise ratio. Especially, faint emission-line objects can be quite well detected in narrow bands containing the line. Therefore, object search is done on the sumframe of each band with SExtractor software \cite{BA96} and the filter-specific object lists are then merged into a master list containing all objects exceeding a minimum S/N ratio on any of the bands. For merging all objects are considered identical which fall into a common error circle of $1\arcsec$, while the typical seeing is $1\farcs5$. The positions of all detections in the different color bands are then averaged into a final position. 

Although the object morphology is determined by SExtractor on each sumframe, we use our own morphological analysis based on MPIAPHOT. The final morphology of an object is determined on the sumframe where the object shows up with the highest S/N ratio. After comparison with the typical PSF of stars, objects are sorted into the classes {\it stellar} and {\it extended}.

The multi-color classification is based on color indices which should be measured as accurate as possible. Therefore, an accurate relative calibration of the different filters is very important. The absolute calibration is not relevant for the calibration, but matters for flux-limited counts or luminosity-dependent studies. For statistically correct results of the classification we need not only accurate flux data but also accurate individual errors for them (see paper I, Sect. 2). We achieve this by measuring the object fluxes on each individual image and deriving the error from the variance of the resulting values. 

We get an optimum signal-to-noise ratio by integrating the photons over an aperture with a Gaussian weight distribution \cite{MR93}. In each image the aperture is located at the same position on the sky and its size and weight distribution is adapted to the seeing of the frame. Every image gets a weight aperture that simulates a Gaussian smoothing to a common seeing before the photon counting, in order to make sure that always the same fraction of an object´s intrinsic light distribution is probed. The obtained flux is  calibrated by the standard stars to yield the accurate flux value for point sources inside a virtual aperture of infinite size. For non-point sources the flux is underestimated by some degree depending, and total fluxes are estimated from morphological parameters. See Meisenheimer et al. 2000 for a detailed description on all aspects of the CADIS data reduction.

For photon counting, the coordinates from the master list are transformed into the coordinate system of each individual frame taking not only translation and rotation into account, but also scaling and distortion differences resulting from the use of different imaging instruments. The measured counts are translated into physical fluxes outside the terrestrial atmosphere by using a set of tertiary spectrophotometric standard stars we established in the CADIS fields. For these standard stars we know the physical fluxes in every CADIS filter, and therefore the calibration of each image is independent of the photometric conditions during the exposure.

\subsection{Color indices, flux units and errors}

Since we present CADIS results in terms of a magnitude scale not commonly used by optical astronomers, we start with a little introduction on basics of magnitude and color systems here. A color index describes an object's brightness ratio in any two chosen filters and is usually given in units of magnitudes. There are several definitions for the zeropoints of the magnitude scale, an astronomical definition handed down from history and an increasing number of modern physical definitions (e.g. in {\it cgs}-units). The zeropoint of the astronomical magnitude scale was for a long time the A0V star Vega = $\alpha$ Lyr, so an object's magnitude was defined for any filter as
\begin{eqnarray}
	m_{obj} = -2.5 \log (F_{obj}/F_{Vega}) ~,
\end{eqnarray}

and astronomical color indices between two filters $A$ and $B$ were given by
\begin{eqnarray}
	m_A - m_B = -2.5 \log \frac{F_{obj,A}}{F_{obj,B}} 
			+2.5 \log \frac{F_{Vega,A}}{F_{Vega,B}} ~.
\end{eqnarray}

The issue of brightness calibration became a little more complex by improvements in detector systems. Different filter systems were used with different detector types and introduced certain inconsistencies in the calibration. Nowadays, Vega is not defining the zeropoint anymore but remains rather close to it with small filter-dependent deviations of $|\Delta m| \la 0\fm05$. However, since the Vega spectrum has a highly non-trivial shape on a physical flux scale, values of astronomical color indices do not  properly convey the related physical flux ratios. The first physical magnitude scale was introduced by Oke (1964, 1965, 1974) and called ``AB magnitude''. It is defined as:
\begin{eqnarray}
	& AB\mbox{mag} = & -2.5\log {F_\nu} - 48\fm60 \\
  & & \mbox{with } ~ F_\nu ~ \mbox{in erg\,cm$^{-2}$\,s$^{-1}$\,Hz$^{-1}$} \nonumber ~,
\end{eqnarray}

so objects with $F_\nu = \rm{const}$ have all their AB colors equal to zero. The denomination ``AB'' originates from a program variable in J. B. Oke's reduction software for the calibration of his standard stars \cite{Oke98}. In the recent past, the ``ST magnitude'' was defined as \cite{Wal95}:
\begin{eqnarray}
	& ST\mbox{mag} = & -2.5\log {F_\lambda} - 21\fm10 \\
  & & \mbox{with} ~ F_\lambda ~ \mbox{in erg\,cm$^{-2}$\,s$^{-1}$\,\AA $^{-1}$} \nonumber ~,
\end{eqnarray}

so objects with $F_\lambda = \rm{const}$ have their ST colors equal to zero, where ``ST'' means probably ``Space Telescope''. In CADIS flux values are given in units of photons per (m$^2$\,s\,nm) as it is done in X-ray astronomy. Therefore, we feel inclined to introduce a third physical system, the ``CD magnitude'', defined as
\begin{eqnarray}
	& CD\mbox{mag} = & -2.5\log {F_{phot}} + 20\fm01 \\
  & & \mbox{with} ~ F_{phot} ~ \mbox{in $\gamma$\,m$^{-2}$\,s$^{-1}$\,nm$^{-1}$} \nonumber ~,
\end{eqnarray}

so objects with $F_{phot} = \rm{const}$ have their CD colors equal to zero. The name ``CD'' relates to CADIS and to the diplomatic choice of this magnitude, that is just the average of AB and ST magnitude for any given object. The three flux scales mentioned are related by
\begin{eqnarray}
	\nu F_\nu = hc F_{phot} = \lambda F_\lambda ~.
\end{eqnarray}

\begin{figure*}
\centerline{\psfig{figure=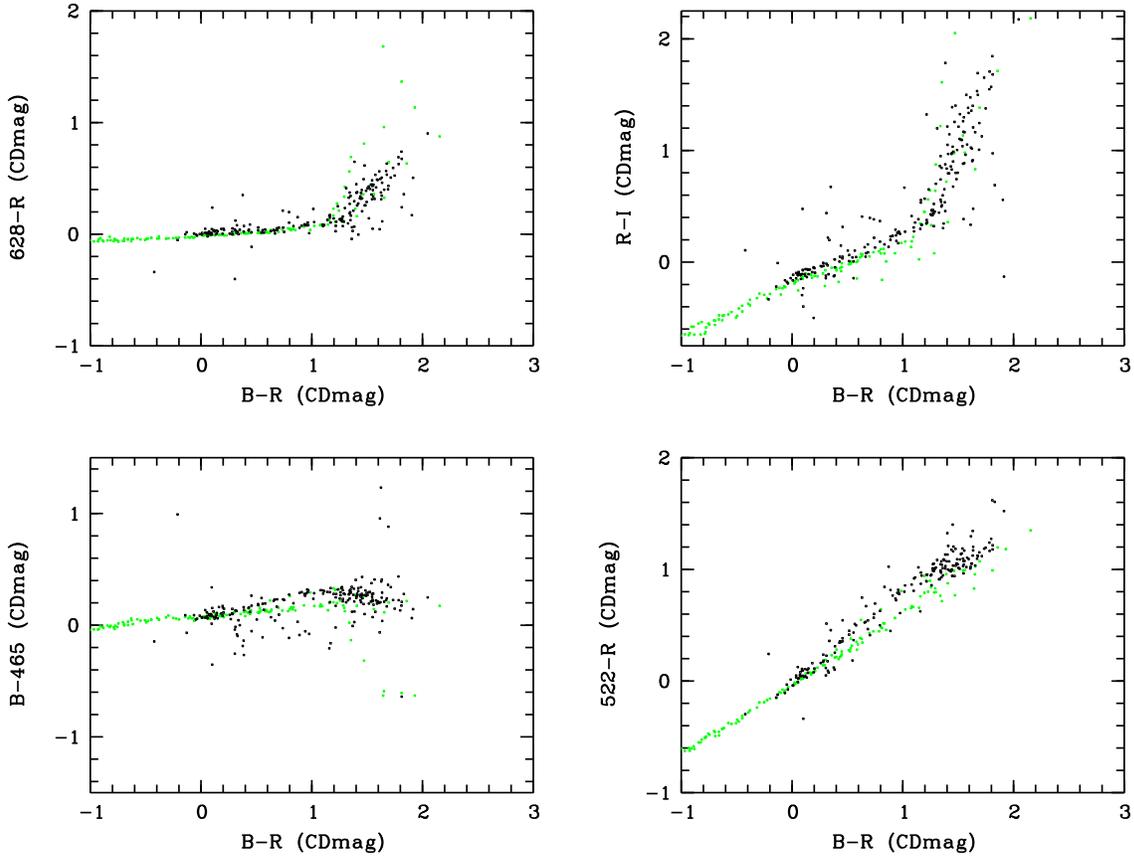,angle=270,clip=t,width=15cm}}
\caption[ ]{This diagram shows a few selected color-color plots containing all point sources with $R<22$ in the CADIS 16h-field of which are roughly 75\% stars, 20\% compact galaxies and 5\% quasars. By plotting the CADIS objects in black and the Pickles (1998) library in grey a number of features can be seen in the diagrams: E.g., the R--I color index in the upper right panel illustrates a calibration offset of $0\fm05$\,mag. The B--465 and 522--R color indices in the lower two panels already contain stellar population information: the CADIS objects (mostly halo stars, population II) fall preferentially onto the upper one of two arms in the color distribution of G and K stars (having $0\la B-R \la 1$), while most library stars (residing in the Galactic disk, population I) are in the lower arm. \label{calibcheck}}
\end{figure*}

All these magnitude systems are in fact designed to have a common zeropoint at a wavelength of $\lambda_0 = 548$\,nm, so every object observed through a quite narrow filter centered there will have
\begin{eqnarray}
	\mbox{$AB$mag = $CD$mag = $ST$mag = astronomical mag}
\end{eqnarray}

For most CADIS objects photon fluxes are given in rather practical units, since an object of $V \approx 20$ has just a flux of 1\,$\gamma$\,m$^{-2}$\,s$^{-1}$\,nm$^{-1}$ at $\lambda_0$, while Vega has almost exactly $10^8\,\gamma$\,m$^{-2}$\,s$^{-1}$\,nm$^{-1}$. As an input to our multi-color classification we define color indices on the basis of CD magnitudes by
\begin{eqnarray}
	m_1-m_2 = -2.5\log\frac{F_{phot,1}}{F_{phot,2}} ~,
\end{eqnarray}

and approximate the corresponding errors for well-detected objects ($>5\sigma$) by
\begin{eqnarray}
	\sigma_{m_1-m_2} = \sqrt{(\sigma_{F_{phot,1}} / F_{phot,1})^2 +
			 (\sigma_{F_{phot,2}} / F_{phot,2})^2} & ~.
\end{eqnarray}

In paper I, we had discussed the relevance of a common {\it base filter} for the various color indices, which is supposed to have relatively small flux errors in order to keep the color errors as low as possible. Since CADIS observes a small number of broad bands and a larger number of medium-band filters (see Fig.\,\ref{filterset}), we decided to form color indices from broad bands neighboring on the wavelength axis, i.e. B--R, R--I and I--J or I--K depending on the availablity of the data. Each of the medium bands we combine with the most nearby broad-band in terms of wavelength, which then serves as a {\it base filter} for the medium-band color indices, e.g. B--486 or 605--R, where letters denote broad bands and numbers represent the central wavelength of medium-band filters measured in nanometers. 

Our I band is in fact a medium-band filter centered at the location of an M star pseudocontinuum feature around 815\,nm. It is among the first filters observed on every field and therefore the most suitable choice for another base filter between R and J. For the classification this means, that we use a few deep broad bands to fit the global shape of the SED, and then use a few groups of medium bands around each deep broad-band to fit the smaller-scale shape locally. This scheme is superior to single fit with all filters, since it can tolerate changes in the global SED due to reddening or cosmic variance, while being more sensitive to distinctive spectral features on a smaller scale.

\begin{figure*}
\centerline{\psfig{figure=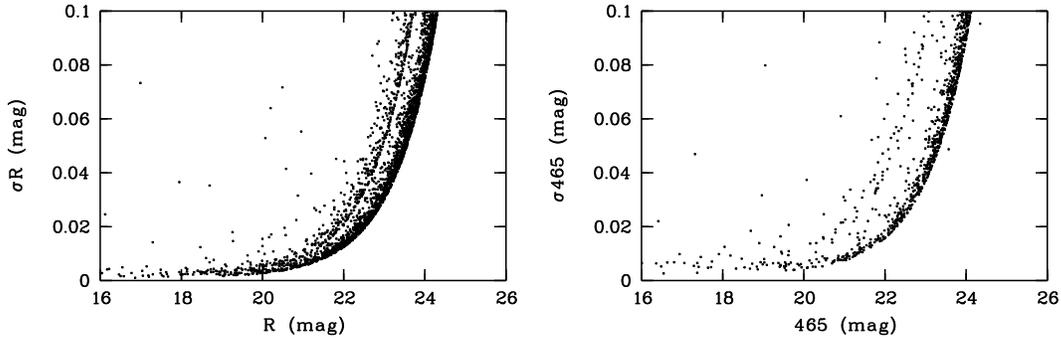,angle=270,clip=t,width=14cm}}
\caption[ ]{This diagram illustrates the flux errors in two example filters and has all objects detected in the 16h-field with an error of less than 10\% in the broad-band filter CADIS-R and the medium-band filter centered at 465\,nm plotted. The sharp lower error limit of parabolic shape is given by pure photon shot noise. Most objects have larger errors due to sub-optimal flatfielding and further reduction effects. The R band shows two well-populated arms since a central square part of the field has been imaged on a smaller CCD before and therefore reaches deeper when all existing images are combined. The thinly populated brighter arm in the 465-image results from objects around the field edges, which are only seen in few exposures due to dithering. The few objects in the upper left area of the diagram have unusually large errors due to uncorrected detector defects or other artifacts affecting one of the involved images. \label{fluxerrors}}
\end{figure*}

Before entering a CADIS object table into the classification procedure, we check the calibration of the color indices by a comparison with the star colors from the Pickles (1998) library. This library is also used for the multi-color classification itself, so its colors are required to be consistent with the star colors in the CADIS data for a successful classification. Since we calibrate our object fluxes by Pickles spectra in the first place, consistent colors are expected unless a fault in the reduction or calibration process altered the flux results. The calibration check involves a visual inspection of the color-color diagrams of all point sources with $R<22$ in a CADIS field (see Fig.\,\ref{calibcheck}), among which 75\% should be stars and the scatter due to flux errors should be small.  Although some faint compact galaxies and quasars are contained among the point sources, the check is still possible since they are only a minority of objects likely to show very different colors. We then ignore these outliers and focus on the bulk of objects, which are almost entirely normal stars.

Any calibration mistake showing up as a zeropoint shift in the color-color diagrams leads to an inspection of the reduction process until the causing fault is found. We aim to detect shifts of $\ga 0\fm03$, which is our goal for the accuracy of the relative calibration. Larger offsets in color indices would limit the performance of the multi-color classification, which is otherwise limited by color differences between real objects and the library as well as by the design of the libraries which itself follows the $0\fm03$-goal. As any calibration of absolute fluxes is probably insecure on the order of 10\% and does not matter for the multi-color classification anyway, there is no independent check for this besides a check for consistency among the tertiary CADIS standards themselves.

This calibration check highlights already the difficulty of disentangling valid astrophysical information from data processing problems. The two diagrams containing the color indices B--465 and 522--R show an apparent disagreement between measurements and library colors for G and K stars which have $0 \la B-R \la 1$, while the agreement is good among F and M stars. In these two mediumband filters different stellar populations form two separate arms: The data are dominated by the halo population which lies above the disk population that makes up most of the library.

\subsection{Photometry and morphology data}

The final CADIS catalog is defined to contain all objects within a circle of $400\arcsec$ radius around the field center, provided they have been detected at least at a 6-$\sigma$ level in any coadded sumframe from a CADIS filter or Fabry-Perot band. These limits have been chosen to avoid spurious effects from the field edges and unnecessary overhead from dealing with noise objects. This round field will contain photometry from almost all filters observed with CAFOS, but those observed with MOSCA will only be available for objects in the central $10\arcmin \times 10\arcmin$. Table\,\ref{exposure_limits} summarizes the filters, exposure times and limiting magnitudes for the three fields analysed here, illustrating on which data the multi-color classification is currently based. These fields have their imaging program completed by 80\% on average and Meisenheimer et al. (2000) reports on the final exposure depths to be reached.

Object photometry is performed on each single frame, so that the scatter among the individual flux measurements gives an estimate for the flux error. Since an object is placed on different locations in the single frames, the error obtained this way includes not only photon shot noise but also effects introduced by an imperfect flatfield or not properly corrected night-sky fringes. If the scatter is unplausibly low due to a chance coincidence of the single values, we still assume a minimum error corresponding to photon noise. Fig.\,\ref{fluxerrors} shows the distribution of flux errors and magnitudes in the broad R band and the medium-band filter at 465\,nm for all objects detected with less than 10\% flux error in the 16h-field.

The two main sources of problems for proper color determination are object variability and object blending. The multi-color database of CADIS is collected over several years, which means that different filters are exposed at different epochs. We aim to identify variable objects by repeating the broad R-band images several times during the survey. This way we have found variability in a fraction of our spectroscopic quasar sample, some variable stars and even a high-redshift supernova candidate with no detectable host galaxy. Object blending is a problem we can not resolve, and for close chance projections we do not even have a possibility to flag objects on the basis of imaging data. In one case we only realized the true nature of a blended object with VLT spectroscopy in $0\farcs5$ seeing. Object blending is furthermore a problem for morphological measurements.

\begin{figure}
\centerline{\psfig{figure=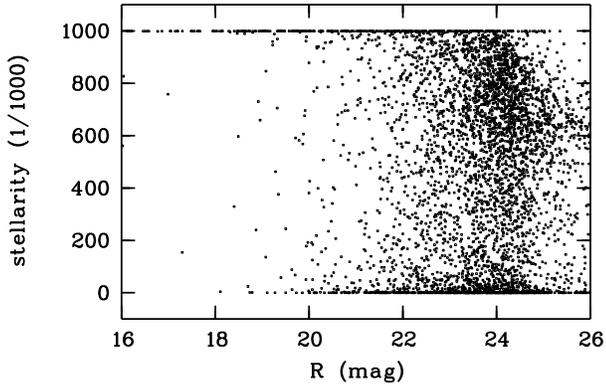,angle=270,clip=t,width=8cm}}
\caption[ ]{Stellarity over R band magnitude for all objects in the 16h-field detected at least at a 6-$\sigma$ level in any CADIS filter or Fabry-Perot band. \label{morph}}
\end{figure}

Fig.\,\ref{morph} shows the morphological properties of all objects in the CADIS 16h-field as determined by MPIAPHOT. Although MPIAPHOT determines the semi-width of major and minor axis as well as an axis angle, all information available is condensed into a stellarity parameter based on the average PSF on a frame. This stellarity parameter ranges from 0 to 1000 with values above 800 being considered as point source images. Since the stellarity value of an object is always determined in the filter where it is detected with the best signal-to-noise ratio, some objects in Fig.\,\ref{morph} are very faint in the R-band but still clearly considered point sources, because they are much brighter in the far-red or even K-band and their morphology is easily determined there. The morphological star-galaxy separation starts failing at $R \ga 21$ where already many galaxies appear compact (stellarity $>800$) in our typical seeing of $1\farcs5$. So, the multi-color classification provides a strong improvement already in this brightness regime, i.e. for most survey objects.

\begin{table}
\caption{Filters and 10-$\sigma$-magnitude limits for the three CADIS fields presented here. Filters not exposed or reduced yet are left blank. Exposure times are given for the 16h-field as an example, and resemble true exposures for the 2.2-m-telescope, or equivalent 2.2-m-exposures in italics if the observations were done at the 3.5-m-telescope. The given magnitudes are astronomical (Vega-normalised) $10\sigma$-limits estimated from the magnitude distribution of those objects with flux errors measured to be roughly 10\%. \label{exposure_limits}}
\begin{tabular}{l|l|lll}
$\lambda_{cen}$/fwhm (nm) & $t_{exp,16h} (s)$ & $m_{1h}$ & $m_{9h}$ & $m_{16h}$ \\
\noalign{\smallskip}
\hline
\noalign{\smallskip}
461/113 (B)  		&  6200 	& 24.6  & 24.8   & 24.7  \\
649/172 (R) 	 	&  5300 	& 24.1  & 24.5   & 24.1  \\
815/25  (I)  		& 30700 	& 21.7  & 22.9   & 22.9  \\
1200/000 (J)	  	&       	&       & 21.5   &       \\
2120/340 (K$^\prime$)   & {\it 22500} 	& 19.9  & 20.0   & 19.5  \\
\noalign{\smallskip}
396/10	 		&       	& 24.0  & 23.4   &       \\
465/10	 		& 38000 	& 23.6  & 23.9   & 24.1  \\
489/20	 		&	 	&       & 23.5   & 23.7  \\
522/15	 		&	 	& 24.5  & 24.3   & 23.4  \\
535/14	 		& 	 	&       & 24.1   & 24.0  \\
611/16  		& 11000 	& 24.1  & 23.4   & 23.4  \\
628/16	 		& {\it 23000}	& 23.5  & 23.6   & 23.6  \\
683/18	 		& 		&       & 23.5   &       \\
702/19			& 		& 23.7  & 23.7   &       \\
752/28			& 15600		& 23.1  & 22.5   & 22.9  \\
855/13			& 		&       & 22.3   & 22.4  \\
909/31			& 31600		& 21.2  & 21.8   & 22.3  \\
\noalign{\smallskip}
\hline
\end{tabular}
\end{table}

\section{Multi-color classification}

Objects are classified independently by their photometric and morphological information, since the morphologic information transmitted through average CADIS seeing ($\sim 1\farcs5$) is not very useful in contrast to the well-resolved spectrophotometric information provided by a dozen filters. A cross check of results turns out to be useful in cases where the CADIS color space does not provide enough discriminative power to distinguish overlapping classes. 

Sect.\,4.1 outlines our classifications scheme, which has been published in full detail in paper I. In Sect.\,4.2 we discuss results from Monte-Carlo simulations of CADIS multi-color data at various magnitude levels, where we like to see what classification performance we can expect from the CADIS filter set. Sect.\,4.3 presents statistical properties of the classified object catalogs for the three fields reported here. Finally, Sect.\,4.4 discusses the fraction of {\it unclassified} and {\it strange} objects, whereas a detailed discussion of a few peculiar objects challenging the classification can be found in the appendix.

\subsection{Classification scheme}

Our multi-color classification scheme essentially compares the observed colors of each object with a color library of known objects. This library is assembled from observed spectra by synthetic photometry performed on the CADIS filter set. As an input we used the stellar library from Pickles (1998), the galaxy template spectra from Kinney et al. (1996) and the QSO template by Francis et al. (1991). From the latter, we generated regular grids of QSO templates ranging in redshift within $0<z<6$ and having various continuum slopes and emission line equivalent widths. Also, a grid of galaxy templates has been generated for $0<z<2$, and contains various spectral types from old populations to starbursts. 

Objects are classified by locating them in color space and comparing the probability for each class to generate the given measurement. Given the photometric error ellipsoid in the n-dimensional color space, each library object can be assigned a probability to cause an observation of the measured colors. For a whole class, this probability is assumed to be the average value of the individual class members, resulting in relative likelihoods for each object to belong to the various classes. Emphasis should be put on the fact, that quasars are selected by a positive criterion ({\it quasar-like objects}) and not by an exclusive rule ({\it unusual objects}). We also note, that we do not use any absolute magnitude information or cosmological knowledge about the typical abundance of different object types. This could be added to the scheme in order to reduce the global rate for failure in the classification, but it would also suppress the likelihood to identify members of the rare quasar class.

This statistical approach contains two implicit concepts of unclassifiability. For each object we normalized the relative likelihoods such, that the three classes would always add up to 100\%. But if the photometric error ellipsoid contains an overlap of several classes, the likelihood is distributed among them and the true class may be {\it unclassifiable}. This occurs when the available colors are not discriminating between the classes or when the measurement error is too large. Unusual objects which are far away from any library in color space, we call {\it strange}, if their measurement is statistically inconsistent with being caused by any library object at more than a 3-$\sigma$ level. 

Since the galaxy and quasar libraries resemble regular grids in redshift and spectral type, these parameters can also be estimated from the observation. For this purpose, we use an advanced Minimum-Error-Variance estimator (MEV+) in order to determine a redshift estimate as well as a an error estimate (see paper I for all relevant details on the multi-color classification and redshift estimation procedure).

\subsection{Monte-Carlo simulations}

\begin{figure*}
\centerline{\psfig{figure=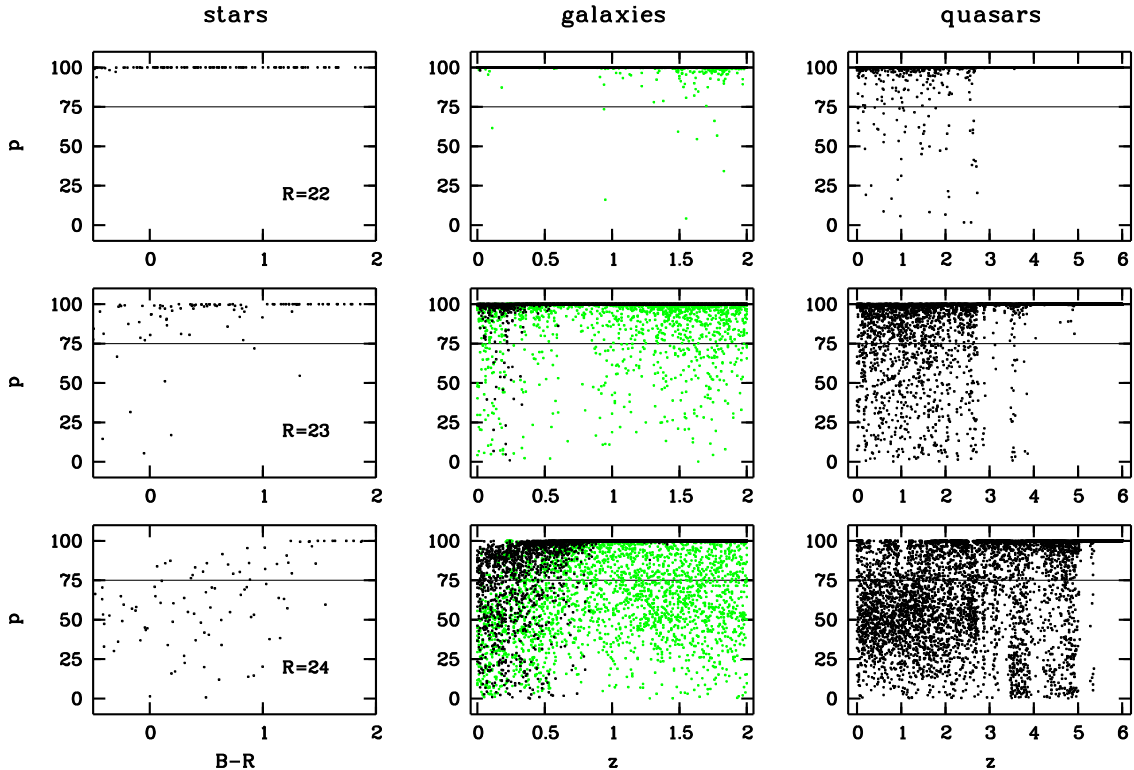,angle=270,clip=t,width=15cm}}
\caption[ ]{Monte-Carlo simulation for the CADIS-classification of stars, galaxies and quasars with $R=22\ldots24$. The probability for a simulated object to be assigned to its original class is plotted over the color $B-R$ for stars and over the redshift for galaxies and quasars. The $B-R$ color is in CD magnitudes, offset by $-0\fm67$ relative to Vega calibrated colors. In case of the galaxies black dots denote quiescent galaxies (SED$<$60) and grey dots are starburst systems (SED$\ge$60). For bright objects the performance is limited by a systematic uncertainty of 3\% assumed as a minimum error for the color indices. \label{3prob}}
\end{figure*}

\begin{figure}
\centerline{\psfig{figure=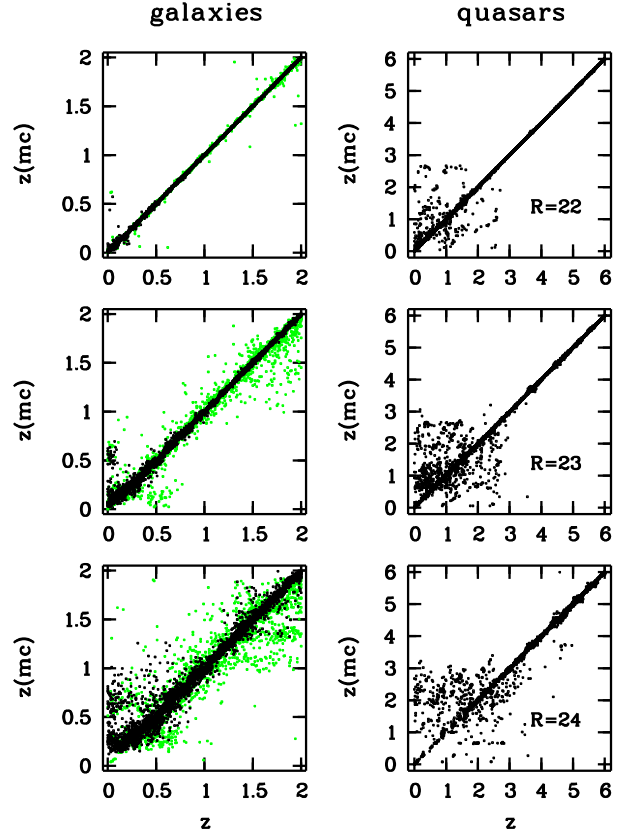,angle=270,clip=t,width=8cm}}
\caption[ ]{Monte-Carlo simulations for the photometric redshifts of galaxies and quasars with $R=22\ldots24$ according to the MEV+ estimator. Among galaxies black dots denote quiescent systems and grey dots are starburst galaxies. This diagram shows the redshift estimates for all galaxies, including the tentative ones, but only for those quasars passing the classification limit of 75\%. \label{2zz}}
\end{figure}

\begin{table}
\caption{Classification matrix for objects of $R=23$ and $R=24$ as derived from Monte-Carlo simulations. An input vector containing a true number distribution of objects among the three object classes would be mapped by this matrix onto a classified distribution among four classes. Numbers below 0.005 are left blank. Due to rounding numbers in one column do not always add up to 1.00. At $R=22$ the three main diagonal elements are essentially all 1.00. \label{class_matrix}}
\begin{tabular}{l|ccc|ccc}
	& \multicolumn{3}{c}{true class, $R=23$} & \multicolumn{3}{c}{true class, $R=24$} \\
{\it classified as} & star	& galaxy	& quasar & star	& galaxy	& quasar \\
\noalign{\smallskip}
\hline
\noalign{\smallskip}
star		& 0.93	& 		& 0.01	& 0.48	& 0.02		& 0.02	\\
galaxy		& 	& 0.95		& 0.01	& 0.04	& 0.70		& 0.02	\\
quasar		& 0.01	& 		& 0.91	& 0.02	& 0.02		& 0.60	\\
unclassified 	& 0.06	& 0.04		& 0.07	& 0.47	& 0.26		& 0.36	\\
\noalign{\smallskip}
\hline
\end{tabular}
\end{table}

\begin{table}
\caption{Mean error of multi-color redshifts as obtained from the Monte-Carlo simulations for the filter set of the 16h-field. \label{tab_sdz}}
\begin{tabular}{l|ccc}
object type		& $R=22$ & $R=23$ & $R=24$ \\
\noalign{\smallskip}
\hline
\noalign{\smallskip}
quiescent galaxy	& 0.01	& 0.05	& 0.13	\\
starburst galaxy	& 0.04	& 0.12	& 0.25	\\
\noalign{\smallskip}
quasar @ $z\le 2.2$	& 0.27	& 0.51	& 0.78	\\
quasar @ $z>2.2$	& 0.10	& 0.17	& 0.22	\\
\noalign{\smallskip}
\hline
\end{tabular}
\end{table}

We carried out a range of Monte-Carlo simulations to check what classification performance we can expect from our method in combination with the CADIS data. We used simulated multi-color observations of stars, galaxies and quasars as an input into our algorithm and compared input and resulting classification. The simulated input list was prepared by using objects from our color libraries.

We assumed a certain R-band magnitude and calculated individual filter fluxes and corresponding errors for each object. Then we scattered the object fluxes according to a normal distribution of the flux errors. Finally, we recalculate resulting color indices and index errors and use this object list as an input to the multi-color classification. From the stars we use just all 131 library members as test objects. From the galaxies we take only every third member of the present library giving us 6700 objects. From the quasar library we use every seventh object resulting in 6450 quasars per test run. The simulations for the quasars are appropriate for surveys measuring the colors wihtin a short period of time since no variability is incorporated skewing color indices in a database collected over a long time as it is the case with CADIS.

These simulations show us how well the classification can possibly work, assuming that real objects will precisely mimic the library objects. Every real situation will contain differences between SED models and SED reality, sometimes called ``cosmic variance'', which will worsen the performance of every real application. Nevertheless, the simulation highlights the principal shortcomings of the method and the chosen filter set in particular. 

We run these tests for stars, galaxies and quasars with magnitudes of $R=$22, 23 and 24, in order to see how the classification performance degrades from optimum to useless with decreasing object flux. We expect that the classification shows its best possible performance already at $R=22$, where the calibration uncertainty of 3\% and cosmic variance will dominate over the photon noise. Finally, at $R=24$ we do not expect the classification to be reliable anymore, but we would like to see how well the redshift estimation still works. Afterwards, we would like to compare these simulations with the real performance derived from spectroscopic identifications, most of which were obtained on the 16h-field. For a fair comparison, the simulation uses just the filter set presently available on this field, which is lacking the four filters 396/10, 683/18, 702/19 and J for full performance.

As a result we obtain a classification and potentially a redshift estimation for every simulated object. The sample results can be condensed into a classification matrix showing what fraction of an input sample from a given class is classified into the various output classes, and especially what objects are seen as {\it unclassifiable}. Table\,\ref{class_matrix} shows the resulting matrix, which for bright objects ($R=22$) basically resembles an identity map corresponding to a classification without mistakes. The main diagonal elements contain the completeness of the three classes, while the other elements count the misclassifications. 

At $R=23$, we expect to loose roughly 10\% of the stars and quasars to unclassifiability. At $R=24$, the survey has become almost useless, since about half of the stars and quasars are not classified correctly anymore. Most incorrectly classified objects are {\it unclassifiable} and only a minority of them is scattered into another class. 

Especially, quasars seem to be not strongly contaminated by false candidates. Only at $R>22$ the contamination should matter. According to the simulations roughly 0.5\% of the galaxies at $R=23$ and 2\% of galaxies at $R=24$ are scattered into the quasar class. If, e.g., the ratio of quasars to galaxies was an the order of 100:1, the contaminants should amount to about a third of the candidates at $R=23$ and should be the dominant fraction at $R=24$.

Faint objects suffering from {\it unclassifiability} have usually rather featureless continua. Fig.\,\ref{3prob} shows a few plots for different classes and R-band magnitudes, illustrating the probability for recovering the correct class of a simulated object in dependence of a characteristic parameter, which is color for stars and redshift for extragalactic objects. The faint {\it unclassified} objects include most stars except for M stars, which have rather characteristic spectra and are bright in the far-red wavelength range, where data from many medium-band filters is available. Starburst galaxies at higher redshift ($z>1$) and quasars at low-redshift ($z<2.2$) show both rather blue spectra, and are not differentiated at the faintest levels. Also, galaxies around zero redshift are not easily recovered, since the CADIS filter set does not contain a U-band or mediumband filters bluewards of 465\,nm. 

The quality of the redshift estimation for galaxies and quasars is shown in Fig.\,\ref{2zz}, where the population on the diagonal are correctly estimated objects and the deviations are mistakes. Most objects scatter around the diagonal and few deviate by a large amount in a non-Gaussian distribution which are {\it catastrophic mistakes}, where the classification considers more than one redshift value as likely, and decides for the wrong one. The variance of the true estimation error obtained from the simulations is summarized in Table\,\ref{tab_sdz}. Generally, quiescent galaxies are estimated more accurately than starburst galaxies due to stronger continuum features, which mainly are the red continuum shape on the blue side of the 4000\,\AA-break. Therefore, also the limiting magnitude for a given redshift accuracy is about one magnitude deeper for the quiescent galaxies. For a similar reason quasars at $z>2.2$ work better than those at lower redshift, since they display a distinctive continuum step across the Lyman-$\alpha$ line.

\subsection{Classification statistics}

\begin{table}
\caption{Classification statistics for the 16h-field. \label{class_tab}}
\begin{tabular}{l|rrrr|rrrr}
        & \multicolumn{4}{c}{stellar} & \multicolumn{4}{c}{extended} \\ 
R-mag   & star & gal & qso & uncl. & star & gal & qso & uncl. \\
\noalign{\smallskip}
\hline
\noalign{\smallskip}
$18\ldots 19$ & 30 &   0 &  1 &   0 &  0 &   9 &  0 &   0 \\
$19\ldots 20$ & 48 &   2 &  1 &   0 &  1 &  23 &  1 &   0 \\
$20\ldots 21$ & 45 &   8 &  3 &   0 &  3 &  81 &  0 &   2 \\
$21\ldots 22$ & 66 &  26 &  4 &   3 & 19 & 177 &  4 &   6 \\
$22\ldots 23$ & 87 &  89 & 11 &  10 & 71 & 347 &  6 &  38 \\
$23\ldots 24$ & 52 & 160 &  7 & 108 & 88 & 493 & 47 & 362 \\
\noalign{\smallskip}
\hline
\end{tabular}
\end{table}

We applied the  multi-color classification to the object catalogs of the fields 1h, 9h and 16h. Although the imaging data collected so far differ in terms of depth and available filters, we expect the classification performance to be roughly similar and use the classification results for various scientific applications. Statistical properties of the classified object samples are demonstrated in the following figures and tables: 

Fig.\,\ref{class_histo} shows magnitude histograms of all objects contained in each of the three fields and the fraction classified successfully  by color, which means that these objects focus at least 75\% of their class membership probability onto a single class. At $R<22$, the  multi-color classification is more than 97\% complete. Towards fainter magnitudes, first the classification becomes incomplete and then even fainter the object counts as well. Tentatively, we consider unclassified objects to be galaxies, simply because they are expected to be the dominant population at faint levels in extragalactic surveys. Fig.\,\ref{class_histo} also shows histograms of all galaxies in each field and the fraction that received an MEV+ redshift estimate. The subsample of galaxies with MEV+ estimates is at least 97\% complete at $R<23$.

\begin{figure*}
\centerline{\psfig{figure=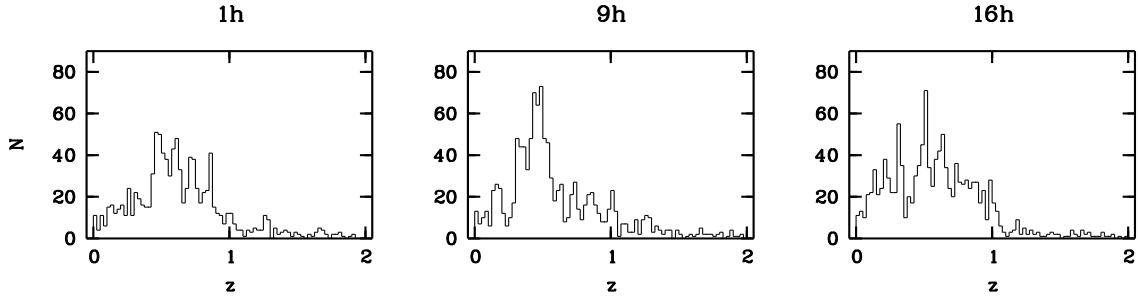,angle=270,clip=t,width=15cm}}
\caption[ ]{These histograms show the redshift distribution of all galaxies with $R<23.5$ and a successful MEV estimate for each field. The galaxy library does not contain objects at redshifts of $z>2$, which is therefore excluded as an estimate by definition. \label{z_histo}}
\end{figure*}

Fig.\,\ref{z_histo} shows galaxy redshift histograms for each field, where the visible features are mostly reflecting cosmic large-scale structure. Of course, the true distribution is smoothed  and potentially distorted slightly by the estimation errors. Also, objects at $z>2$ are excluded by definition of the galaxy library. 

Table\,\ref{class_tab} lists the  distribution of multi-color classes over magnitude and morphology for the 16h-field, where the contribution of compact galaxies can be seen as well as certain numbers of seemingly extended faint stars. There are two reasons for these apparent contradictions: First, there is truely a population of galaxies not resolved in our seeing and also a small fraction of stars that appear extended because of chance projections with fainter extended objects or because they are double stars. Object blending is of course also a problem for color determination and classification. Second, also the morphology information degenerates towards fainter magnitudes as does the color classification, due to PSF variations and photon noise. As demonstrated in Fig.\,\ref{morph}, the morphology information has become useless for not clearly extended sources at $R\sim 24$.

\begin{figure*}
\centerline{\psfig{figure=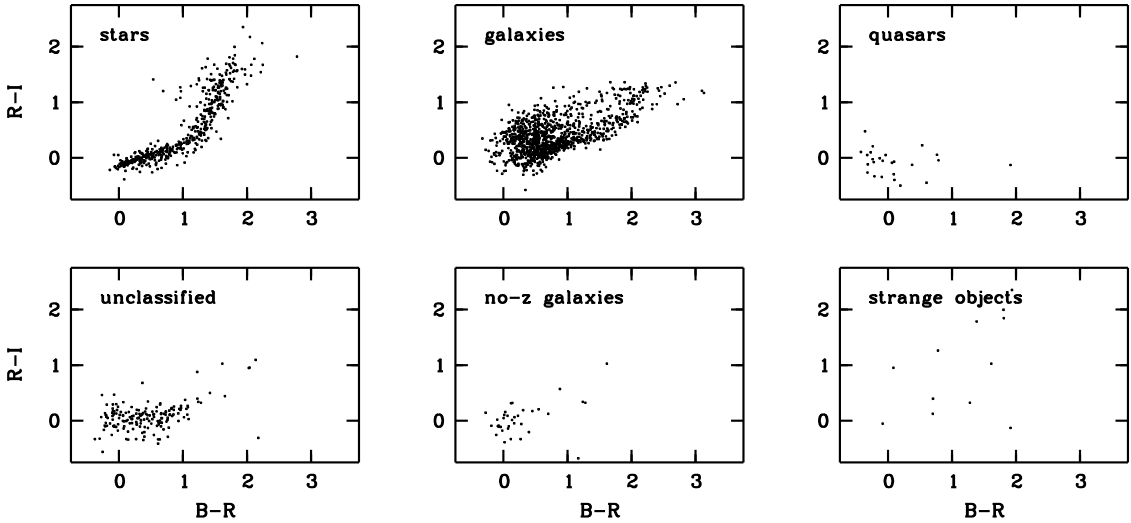,angle=270,clip=t,width=15cm}}
\caption[ ]{These B--R vs. R--I color-diagrams illustrate the location of all 1785 objects with $R<23.5$ in the 16h-field separated in the following groups: The panels in the top row show the class populations after final classification. The panels in the bottom row show unclassified objects, i.e. tentative galaxies (left panel), galaxies without a successful MEV redshift estimate (center panel) and {\it strange} objects (right panel). Color indices are in units of CD magnitudes, where Vega has $B-R=-0.67$ and $R-I=-0.53$. \label{bri6}}
\end{figure*}

Fig.\,\ref{bri6} illustrates the location of different object classes in color-color diagrams always showing B--R vs. R--I. Obviously, the main classes {\it stars}, {\it galaxies} and {\it quasars} share common regions, which they do in any two-color diagram since there is no possible choice of three filters that separates the three classes unambigously. Since our classification takes all multi-color information simultaneously into account, we can still distinguish the classes. 

The bottom row of panels in Fig.\,\ref{bri6} depicts unclassified objects on the left, most of which are supposedly faint blue galaxies overlapping in color space with low-redshift quasars and bluer stars. The center panel contains galaxies without an MEV+ redshift estimate, half of which are also unclassified  by the multi-color algorithm. These objects are at the faint end of the magnitude range used and lie in a region of color space where blue galaxies of quite different redshift clump so closely together, that the typical photon noise of faint objects allows no clear redshift estimation any more. 

Finally, the right panel shows the {\it strange objects}, defined to be inconsistent with any library member at least on a 3-$\sigma$ level. They amount to $\sim 1$\% of the entire object catalog with $R<23.5$ in the 16h-field (see Sect.\,\ref{unclstrange} for details). The level of inconsistency for a truely {\it strange} object is of course determined by its brightness via the magnitude errors. At faint levels like $R\approx 24$, we would require a very strange object indeed to still notice it.

\begin{figure*}
\centerline{\psfig{figure=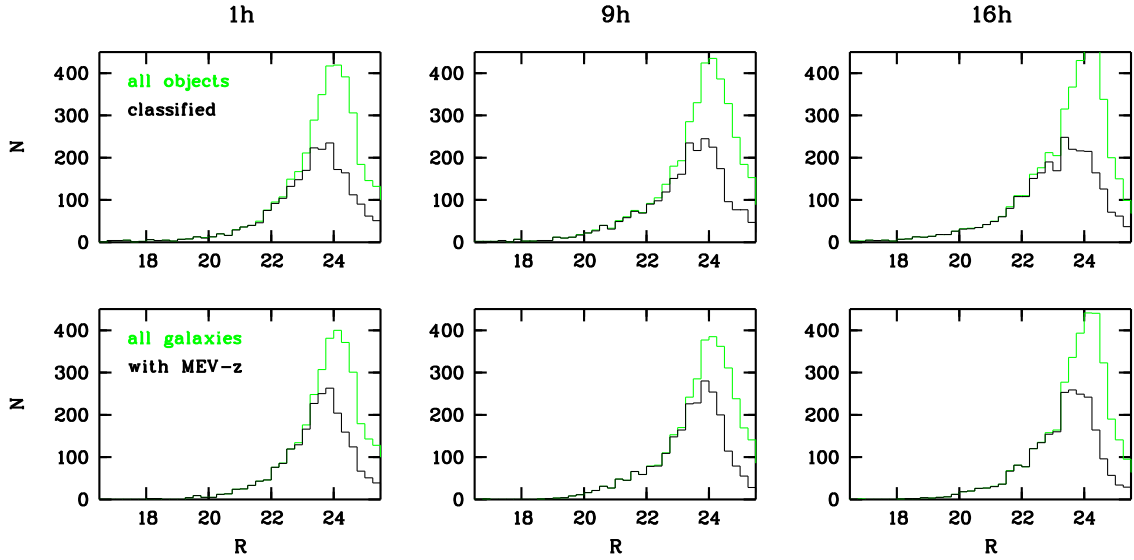,angle=270,clip=t,width=15cm}}
\caption[ ]{The panels in the top row show number histograms for all objects in individual field catalogs (grey line) and for the subsample of objects classified by concentrating more than 75\% probability on a single class (black line). The remaining objects are tentatively classified as galaxies (on purely statistical grounds). The panels in the bottom row show which part of all galaxies (grey line, including tentative galaxies) get a successful MEV redshift estimation (black line). The remaining galaxies have an estimated redshift error of $\sigma_z \ge 0.25$.
\label{class_histo}}
\end{figure*}

\subsection{Unclassified and strange objects}\label{unclstrange}

To investigate the issue of unclassifiability further, we took a closer look at the unclassified objects with $R<22$ in the 16h-field. These are 12 objects containing six {\it artifacts}, one blended double source with $2\arcsec$ separation and only five {\it intrinsically unclassified} objects. The six {\it artifacts} should ideally not appear in the catalog at all, because they are close to the edge of the field in one filter which has probably altered their photometry, or because they are second detections of a bright source with different position and photometry. In fact, data artifacts blow up the catalog sample of unclassified objects to some extent and explain part of the difference observed between the Monte-Carlo simulations and the real data.

The five {\it intrinsically unclassified} objects amount to less than 1\% of the catalog sample with $R<22$. They are all extended, not {\it strange} and show galaxy probabilities ranging from 40\% to 75\% with the remaining probability assigned to the star class. One object is a galaxy with $15\arcsec$ visible diameter that is supposedly at extremely low redshift ($z<0.05$). The other four galaxies are estimated to be in the range of $z=0.2\ldots 0.4$, which has some overlap with stars as we know from the faint spectroscopic sample. Three objects lack data for the filters observed with MOSCA, since they lie in the outer range of the round CAFOS field not covered by the squared MOSCA field. Obviously, with fewer filters the classification reaches less deep. We summarize, that the presence of the observed unclassified objects can be explained and is consistent with the expectations from the simulations.

We also investigated the {\it strange} objects with $R<24$ in the 16h-field. Among these 24 objects are nine artifacts and 15 {\it intrinsically strange} objects, corresponding to $\sim 0.6\%$ among a catalog of 2582 objects with $R<24$ in total. The 15 objects are:
\begin{itemize}
  \item  nine rather obvious M stars with colors deviating from the not entirely representative library which does not cover the whole natural spread in metallicity
  \item  three spectroscopically confirmed quasars, but quasar colors tend to look {\it strange} anyway due to variability, variations in emission-line strength, or diverse and odd spectral characteristics commonly observed but not covered by the library
  \item  two $z\approx1$-galaxies combining a very red 4000\,\AA-break around 800\,nm with a rather flat spectrum bluewards. 
  \item  one later M star with an unusually blue B--R color index of 0.53 in CD magnitudes ($=1.2$ on a Vega normalised scale)
\end{itemize}

Only the last three objects are not explained by systematic effects and could well be chance projections of a faint blue object with a brighter red one. But indeed, a Gaussian distribution should naturally contain 0.25\% objects exceeding the 3-$\sigma$ level of {\it strangeness} corresponding to six objects in the $R<24$ sample of the 16h-field. The other two fields also do not contain unusual numbers of strange objects. Therefore, it seems that hardly any exciting unusual objects are in sight which would call for a fundamentally new explanation. On the other hand, the classification quality benefits from the good agreement between the observations and the library.

\section{Spectroscopic check of the classification}

\subsection{Spectroscopic observations}

The quality of the classification and redshift estimation was checked by multi-slit spectroscopy in two dedicated observing runs with MOSCA at the 3.5-m-telescope. In July 1997, three multi-slit masks were observed on the 16h-field with grism green-500 for 4000\,s, 12000\,s and 16000\,s, respectively, yielding 61 identifications of stars, galaxies and quasars. In October 1997, only six more objects received identifications during observations of two masks under bad observing conditions. The objects were selected to have mostly $R<22$ and represent the range of classes, redshifts and colors found in the catalog, except for the quasars, where all candidates with $R<22$ were checked. A few fainter objects were included where they fitted into the slit arrangement of the masks.

Advantage was taken of further CADIS spectrocopy launched for confirmation of emission-line galaxies, by placing additional objects on otherwise empty spaces of the multi-slit masks during three MOSCA runs in January 1998, January 1999 and April 2000, two runs at the Keck telescope in director's discretionary time in June 1997 and January 1998 and one run at VLT in November 1999. Also, some longslit spectroscopy was done on bright galaxies as a backup program to CADIS when the seeing was above $2\arcsec$. Altogether, it was possible to collect 95 more identifications in the three fields reported, yielding a total of 162 identified objects.

\begin{figure*}
\centerline{\psfig{figure=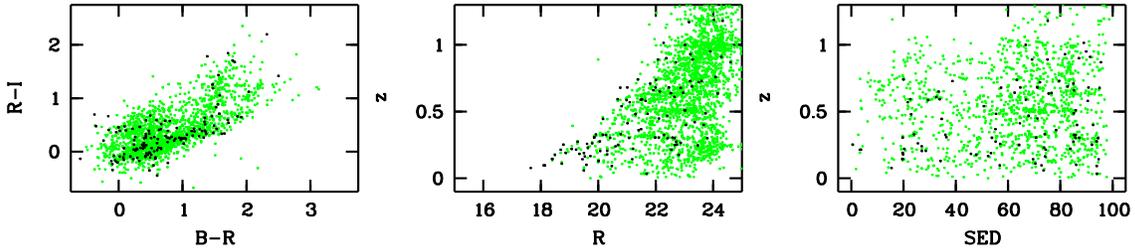,angle=270,clip=t,width=15cm}}
\caption[ ]{These diagrams show various properties of the spectroscopic subsample (black dots, from all three fields) among all objects (grey dots, only from the 16h-field) in the left panel and among the respective galaxies in the center and right panel. The left panel shows just a color-color plot, the center panel plots redshift over R-magnitude and the right panel redshift over galaxy SED. In the left and right panel, a magnitude limit of $R<23.5$ has been used.  \label{specsel}}
\end{figure*}

This subsample of 162 identifications is more or less representative for the object catalog as a whole as illustrated in Fig.\,\ref{specsel}. In terms of galaxy SEDs the spectroscopic sample is not entirely representative, since at redshift $z>0.7$ it contains no red galaxies as opposed to the whole CADIS sample. We note, that red galaxies are expected to receive more accurate redshift estimates due to a stronger 4000\,\AA-break, and therefore the whole CADIS sample should perform equal or better than the spectroscopic subsample in redshift estimation.

\begin{figure*}
\centerline{\psfig{figure=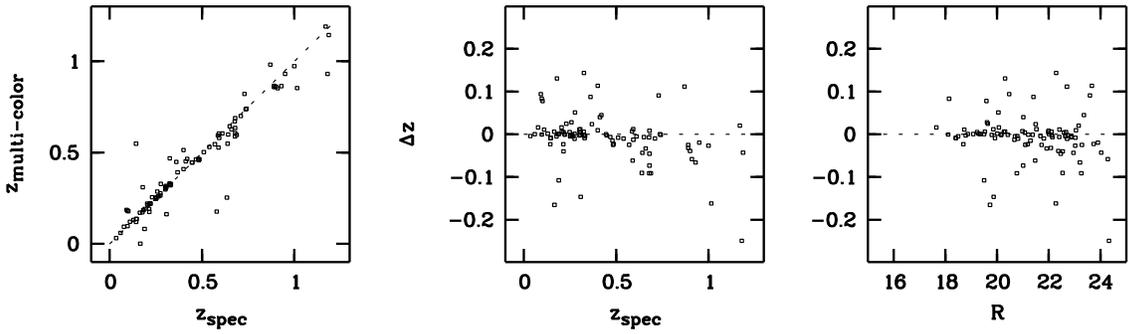,angle=270,clip=t,width=15cm}}
\caption[ ]{The estimation quality for the redshift in the spectroscopic galaxy sample is shown in these diagrams. The left panel plots the multi-color redshift vs. spectroscopic redshift with the highest redshift galaxies residing at $z \approx 1.2$. The center panel shows the error of the estimate ($\Delta z = z_{multi-color}-z_{spectroscopic}$) over redshift, and in the right panel $\Delta z$ is plotted over the R-band. Half of the galaxies are estimated within an error margin of $\pm0.02$.  \label{specgal}}
\end{figure*}

\begin{table}
\caption{Classification matrix for the spectroscopic subsample of 162 CADIS objects, where top rows are objects with $17<R<22$ and bottom rows are objects with $R>22$ (on average $R=23$). \label{spec_matrix}}
\begin{tabular}{l|ccc|ccc}
classified as	& \multicolumn{3}{c}{stellar objects} & \multicolumn{3}{c}{extended objects} \\
		& \multicolumn{3}{c}{true class} & \multicolumn{3}{c}{true class} \\
$17<R<22$	& star	& galaxy	& quasar & star	& galaxy	& quasar \\
\noalign{\smallskip}
\hline
\noalign{\smallskip}
star		& 22	& 		&  	& 1 	&  		&  	\\
galaxy		& 	& 4		& 2	&  	& 58 		&  	\\
quasar		&  	& 		& 15	&  	&  		& 1 	\\
unclassified 	&  	& 		&  	&  	&  		&  	\\
\noalign{\smallskip}
\hline
\noalign{\smallskip}
$R>22$		&	&		&  	&	& 		&	\\	\noalign{\smallskip}
\hline
\noalign{\smallskip}
star		& 5 	& 1		&  	& 1	& 4		&  	\\
galaxy		& 1	& 5		&  	&  	& 25		&  	\\
quasar		&  	& 5		& 2	&  	& 4		&  	\\
unclassified 	&  	& 2		&  	&  	& 3		& 1	\\
\noalign{\smallskip}
\hline
\end{tabular}
\end{table}

\subsection{Quality of the classification}

Table\,\ref{spec_matrix} shows the classification matrix derived from the spectroscopic cross check, separated into 103 bright objects ($17<R<22$) and 59 faint objects ($R>22$, including 11 objects with $R>24$), as well as separated into 64 stellar and 98 extended sources to give also a cross check between morphological appearance and spectroscopic class. The table resembles effectively a 4-D matrix and allows the following conclusions:
\begin{itemize}
  \item  The bright sample contains only two misclassified objects among 103 in total, which translates into $\sim 98\%$ correct classifications. The mistakes are Seyfert-1 galaxies (i.e. quasars) found by chance among the compact galaxies. 
  \item  The faint sample contains 25\% misclassifications and 10\% unclassified objects, with most of them being galaxies. The others are one L star with $R>26$ and $K^\prime=18.5$ and one Seyfert-1 galaxy with $R=22.9$ and $z=1.40$.
  \item  The presence of many compact galaxies in the faint sample was confirmed emphasizing the superiority of a multi-color classification with respect to the morphological analysis. 
  \item  The faint sample shows quite a few galaxies contaminating the multi-color sample of quasars, while the bright sample works very fine. It remains to be shown around what magnitude level in the range $R=22\ldots 24$ the contamination by galaxies becomes critical.
  \item  Based on our small sample of six objects, our default interpretation for unclassified objects as being galaxies appears reasonable.
  \item  Extended objects are almost all galaxies. While in the bright sample one apparently extended looking star and one quasar are correctly classified by their colors, there appears some confusion in the faint sample.
\end{itemize}

There are two extended looking objects in the bright sample, which are correctly classified as a star and a quasar, respectively. The former is in fact a double star at roughly $1\farcs5$ separation with $R=17.3$, and the latter is a low-luminosity quasar at $z=1.57$, where we supposedly resolved a bright host galaxy. The one extended looking star in the faint sample is an M star of $R=25$, where we do not expect a successful morphological analysis. Another extended looking quasar is in fact a Seyfert-1 galaxy at $z=1.4$ and $R=22.9$. The spectroscopic sample contains eight more extended faint galaxies wrongly classified as stars or quasars. 

Among seven faint compact quasar candidates, we confirmed only two, and among four faint extended candidates none, suggesting that we should use a morphological preselection in the faint sample where galaxy contamination appears to take over. But the classification of faint objects has changed in the past while the multi-color database has grown. Also, morphological types may have changed with the addition of deeper images or such of better seeing. Given that the faint sample is still rather small, we can not train our classification to combine color and morphology into a robust procedure at this point. 

Altogether, the classification is close to ideal at $R<22$, but at fainter levels, the abundant galaxies start contaminating the star class and the quasar class. In fact, the best performance is presently achieved by adding a morphological criterion to the classification that assigns automatically the galaxy class to spatially resolved faint objects with $R>22.5$. In this scheme the classification produces 6 errors among 151 spectroscopic identifications with $R<24$.

The contaminants to the star class are normal starburst galaxies at $z=0.25\ldots 0.4$ crossing the stellar locus around G type stars. Contaminants to the quasar class are partly quiescent galaxies at $z\approx 0.3$ which cross the quasar locus around $z\approx 3.5$, since the 4000\,\AA-break of the galaxies mimicks the continuum step over the Lyman-$\alpha$ line. A solution to this problem might be an inclusion of more mediumband filters bluewards of 500\,nm. A second confusion arises from strong starburst system around $z\approx1.2$ crossing the quasar locus around $z\approx 1\ldots 2$.

In the Monte-Carlo simulations (see Sect.\,4.2) for $R=22$ the classification appeared equally ideal as in the spectroscopic cross check. Also, the simulations for $R=23$ and $R=24$ suggest shortcomings of several kinds for the classification of fainter objects, some of which have been observed in the cross check. E.g., the known confusion of quasars at $z\approx 3.5$ with other objects is seen in Fig.\,\ref{3prob} as the line of dots reaching down to zero probability. 

The abundance of low-redshift galaxies sharing their region in color space with quasars causes many more galaxies to show up in the quasar class than vice versa. The 16h-field seems to contain $\sim 500$ galaxies in the magnitude bin of $22<R<23$ and $\sim 1200$ galaxies in the next fainter bin of $23<R<24$. According to the simulations, we expect 0.4\% of the galaxies in the brighter bin and 1.5\% of the galaxies in the fainter bin to scatter into the quasar candidate list, which amounts to $\sim 2$ objects and $\sim 20$, respectively.

Still it seems, that the Monte-Carlo simulations are underestimating the fraction of mistakes in the classification a little. E.g., the fraction of unclassified objects in the real catalog sample of the 16h-field is around 10\% for $R=22\ldots 23$ and around 40\% for $R=23\ldots 24$, while simulations for these intervals yield $\sim 5\%$ and $\sim 20\%$, respectively. As a rule of thumb the simulations might be too optimistic by up to half a magnitude. The dominant reason for this is supposedly cosmic variance, i.e. differences between library and real spectra which exceed the tolerance given by photon noise and calibration inaccuracies. In addition, quasars suffer from variablity during the observational 5-year period of CADIS.

We conclude, that roughly down to $R=22.5\ldots 23.0$, the classified object catalogs can be analysed without any knowledge of the more subtle features in the data. Statistical studies going fainter than this have to worry about completeness issues and about contamination of star and quasar samples by certain types of galaxies. But given the large number of faint galaxies, any loss of galaxies to the star and quasar class is unlikely to be significant for statistical studies on the galaxy sample.

\subsection{Quality of the redshift estimation}

\begin{figure}
\centerline{\psfig{figure=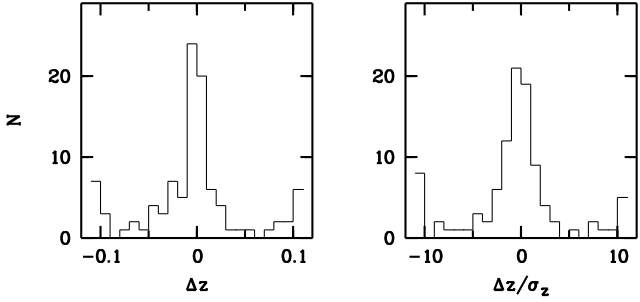,angle=270,clip=t,width=8.5cm}}
\caption[ ]{Most galaxy redshifts are estimated with a $\Delta z$ error variance of $\sim 0.03$, but $\sim 10\%$ of the galaxies receive completely wrong redshift assignments with $\Delta z > 0.1$  (left). For 70\% of the galaxies the true error distribution matches up with the one expected from the multi-color errors, but 30\% of the objects have true errors larger than the estimated 3-$\sigma$-errors  (right), which are mostly starburst galaxies. The reason for the increased scatter in general is, that the observed SEDs are not perfectly matched by the library SEDs. \label{dzhist}}
\end{figure}

The performance of the redshift estimation among the 101 galaxies with spectroscopic redshifts and MEV+ redshifts is illustrated in Fig.\,\ref{specgal} and Fig.\,\ref{dzhist}. Here, four galaxies blended into two double objects have been removed from the sample, since their strongly differing redshifts (0.616 and 1.20 mixed in one pair, 0.24 and 0.729 mixed in the other pair) cause mixed colors unlikely to be useful. Within the given magnitude range of $17<R<24.5$ just above 10\% of all galaxies have large redshift errors $|\Delta z| > 0.1$ which we call {\it catastrophic mistakes}. Examining the remaining sample leads to a distribution with zero mean error and $\sigma_z \approx 0.03$ rms deviation. This result was achieved without any color adjustment of the galaxy library to the real data. Among nearby galaxies catastrophic mistakes tend to overestimate the redshift while those at higher redshift are rather underestimations.

The redshift estimation for galaxies compares also rather well between simulations and real data, except for the few catastrophic mistakes which happen even at brighter magnitudes in the saturation range of the estimation quality. Our interpretation is again that cosmic variance plays the major role in this phenomenon. If these mistakes among bright objects are ignored, the redshift accuracy achieved matches perfectly the simulated results from Table\,\ref{tab_sdz}. We note that no spectrum is available for any CADIS galaxy at $z>1.2$, so we are not in a position to check the estimation quality there. The simulations do not point at specific problems, only the general scatter should be large at the rather faint magnitudes expected for these galaxies.

We also compared the true redshift errors $\Delta z$ with the errors $\sigma_z$ estimated by the multi-color technique itself on the basis of photometric errors and the galaxy distribution in the color space. The ratio $\Delta z/ \sigma_z$ evaluates the error consistency of our redshift estimate. If the estimated errors were representative of true errors, this ratio should have a Gaussian distribution with an rms of 1.0. In fact, it turns out that for 30\% of the galaxies this inconsistency is larger than 3-$\sigma$, while the remaining 70\% show a more or less Gaussian distribution with an rms scatter of 1.2 (see Fig.\,\ref{dzhist}). This result implies that for one third of the spectroscopic galaxy sample, the redshift estimation process considers itself too accurate, supposedly a consequence of cosmic variance that changes the galaxy SEDs and their estimated redshifts while not changing the photometric errors.

\begin{figure}
\centerline{\psfig{figure=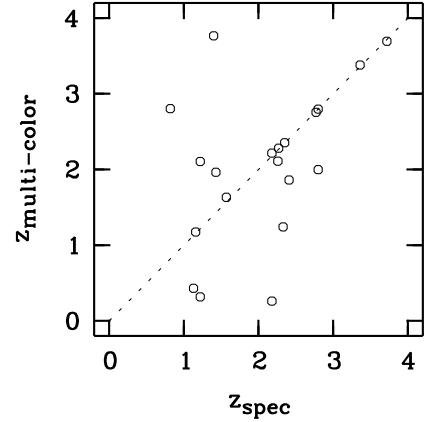,angle=270,clip=t,width=5.5cm}}
\caption[ ]{The estimation quality for the redshift in the spectroscopic quasar sample is shown in this plot of multi-color redshift vs. spectroscopic redshift. The highest redshift quasar resides at $z \approx 3.7$. Half of the quasars are estimated within an error margin on the order of $\pm0.1$, but the other half suffers from {\it catastrophic mistakes}. \label{specqso}}
\end{figure}

Among the 21 spectroscopic quasars only the least luminous and most nearby object has no MEV+ redshift estimate. The performance of the redshift estimation among the remaining 20 objects is demonstrated in Fig.\,\ref{specqso}. Half of the quasars are rather well estimated with a mean redshift error of $+0.01$ and $\sigma_z \approx 0.03$ rms deviation. For the other half of the objects the redshift estimates are completely wrong. The problem here is not only the lack of detectable continuum features for low-redshift quasars, but especially the intrinsic long-term variability of quasars offsetting the magnitudes measured in the various bands depending on the actual epoch of observation.

The simulations for the redshift estimation of quasars at $R=22$ agrees well with the cross check spectroscopy at $z>2$, but it underestimates the error of low-redshift objects. Again, the key problem here most likely is variability that scrambles the real spectrophotometric data collected over several years, while the simulation assumes an instantaneous measurement of colors by taking them from the color library.

\subsection{Potential improvements of the classification}

The classification works essentially well for $\sim 99\%$ of the objects with useful magnitudes. Among the exceptions are some objects which appear {\it strange} to the classifier, although they belong to physically common object classes. Our scheme of classification is based on three fundamental ingredients: {\it data}, {\it library} and {\it classifier}. Improvements to the capabilities of the classified catalog could thus be achieved in the following form:
\begin{enumerate}

 \item Improve the {\it data} by observing the missing bands, or if desired add filters or go deeper. From the missing bands we expect a more accurate redshift estimation for some objects and a slightly smaller fraction of unclassifiable objects. If the data were collected within a shorter period of time, the influence of variability could be reduced. But variable objects are in CADIS anyway detected by repeated R-band observations during many runs.

 \item Improve the {\it library} by, e.g., adding some mixed templates for active galactic nuclei of low luminosity, adding a couple more stars at the low-temperature end, increasing the spread of metallicities among M stars. It is not obvious how to best account for all oddities imaginable among quasars and how to account for their variability when it comes to redshift estimates. Especially, it is not clear whether an enlargement of the quasar parameter space would improve the result. Having the phenomenon not under perfect control an approach as simple as ours might be better.

 \item Improve the {\it classifier} and estimator. This seems to be an issue basically at the faint level, where the number of classification mistakes becomes significant. The mistakes could be reduced, e.g. by taking class richness into account for the membership probabilities (which we do anyway for the unclassified objects by simply taking them all as galaxies), but rare valid quasar candidates could then go lost as most likely galaxies. Eventually, at the limits of the survey performance different applications might require different approaches.

\end{enumerate}

\section{Summary}

In paper I an innovative method for identifying stars, galaxies and quasars in multi-color surveys was presented, which uses a library of $\ga 65000$ color templates for comparison with observed objects. The method aims for extracting the information content of object colors in a statistically correct way, and performs a classification as well as a redshift estimation for galaxies and quasars in a unified approach based on the same probability density functions. 

The three basic ingredients to this method are 
\begin{enumerate}
 \item  accurately measured and calibrated color data for the objects
        to be classified including the color errors
 \item  accurate and representative color template libraries covering
        the range of objects expected in the dataset obtained
 \item  a statistical classifier and parameter estimator which can potentially
        be trimmed for best performance in particular applications.
\end{enumerate}

Also, in paper I it was concluded that medium-band surveys are expected to deliver a performance superior to pure broad-band surveys even under the constraint of equally limited telescope time. Based on survey simulations this method should be capable of
\begin{enumerate}
 \item  separating stars from galaxies down to deeper limits than possible
        by using morphology only
 \item  finding quasars also at redshifts where its colors overlap with the stellar locus
        in two-color projections, particularly in the range of $2.2<z<3.5$
 \item  selecting quasar candidates without reference to their morphology so that objects
        of low-luminosity with resolved hosts would not be excluded from the sample obtained
 \item  using multi-color redshifts from medium-band surveys directly and without
        spectroscopic follow-up for statistical investigations, e.g. for studying
        the evolution of galaxy populations with redshift.
\end{enumerate}

In this paper, we applied the classification scheme to a real multi-color dataset provided by the Calar Alto Deep Imaging Survey (CADIS) investigating its reliability and accuracy. The dataset is 80\% complete on three CADIS fields presented here, the 1h-, the 9h- and 16h-field. Some filters on some fields are still missing and observations of some others have not reached the desired depth, yet. In any case, the CADIS filterset was tailored to the requirements of the emission line survey and is not an optimal choice for a general multi-color classification. 

The libraries of stars, galaxies and quasars presented in paper I are sufficient to classify all but a handful of unusual objects in the CADIS dataset as the fields do not contain significant numbers of objects missing in the libraries. This is demonstrated by the low fraction ($<1\%$) of {\it strange} objects in the sample at $R<24$. Our classification differentiates between {\it unclassified} objects that can not securely be decided among the alternatives and {\it strange} objects that are outliers in color space inconsistent with any library object. We note that unclassified objects are typically not strange and vice versa.

The classified subsamples of stars and galaxies agree with expected numbers and the unusually high content of quasars at $z>2$ was confirmed by spectroscopic observations \cite{Wolf99}. The fraction of unclassified objects is less than 1\% at $R<22$ and reaches about 50\% at $R=24$ due to increasing photon noise that tends to make different original SEDs equally likely sources of the observed colors. At some level in between the classification becomes incomplete and also features an increasing fraction of actual mistakes. These are basically all members of the rich galaxy population spilling over into regions of color space that is usually occupied by stars and quasars.

A spectroscopic cross check using 162 identifications confirmed the multi-color classification to work essentially free of errors at $R<22$ (two mistakes among 103 objects). We do not have proper knowledge about where and how the classification collapses exactly, which is particularly important for the rare quasar class which will become incomplete and dominated by contaminating galaxies at some level. To settle this uncertainty, dedicated spectroscopic observations are required. These would be a valuable and not too time-consuming investment, if a large number of spectra from a CADIS size field could be taken simultaneously with an instrument like VMOS at the VLT.

The findings on the classification performance are consistent with the expectations raised by the Monte-Carlo simulations of the CADIS filter set. They suggest the classification to work nearly perfect down to $R \approx 23$ except for some contamination of quasar candidates by emission line galaxies towards that limit. The simulations actually appear to be too optimistic about the working depth of the classification by about a third of a magnitude. Of course, a real survey will perform worse than a simulation, that can not account for differences between the real world and our library and does not contain artifacts and variable or blended objects. The latter issues have to be dealt with better data analysis.

For galaxies, our multi-color redshifts are useful down to $R \sim 24$. The statistics on the redshift errors are dominated by $\sim 10\%$ catastrophic mistakes, where the estimator decides for the wrong one among alternative values with comparable probability. The core of the error distribution has a zero mean error and $\sigma_z \approx 0.03$ rms width. Quiescent galaxies tend to work better than starburst types.

Half of the quasars receive remarkably correct estimates with an average variance of $\sigma_z \approx 0.1$, preferentially the $z>2$-objects and those of higher luminosity. In contrast, a large amount of redshift confusion is expected at lower redshift. Also, simulations for quasars work better than the real redshift estimation, probably mostly due to variability.  

Eventually, the classified multi-color catalog has been used for several dedicated studies published in separate papers:
\begin{enumerate}

\item The star-galaxy separation with the multi-color data reaches deeper than a morphological separation. Phleps et al. (2000) could therefore use the sample of stars to probe the stellar content and the Galactic structure along the pencil beams established by the CADIS fields, where they find strong evidence of a thick disk.

\item The multi-color redshifts of galaxies are sufficiently accurate for most statistical studies. Also, the classification includes a substantial fraction of compact galaxies into the sample. Using that, Fried et al. (2000) investigated the evolution of galaxies within $0.3<z<1$, finding evidence that {bf strong} evolution takes place only among starburst objects. 

\item The multi-color data allow to identify quasars rather free of contaminants. Spectroscopy is not required to clean a sample obtained by the classification. The selection should also be rather complete, since CADIS finds an unexpectedly high density of $z>2$-quasars. Many of these might reside close to the stellar locus and be overlooked in broad-band surveys \cite{Wolf99}. Thus, quasars can be identified with a much more uniform completeness across the accessible redshift range and more homogeneous samples can be obtained. At $z>2$, it will be possible to constrain the evolution of the luminosity function from a large enough multi-color sample.

\end{enumerate}

\begin{acknowledgements}
The authors thank H.-M. Adorf for helpful discussion about classification methods and their fine-tuning. We also thank the Calar Alto staff for their help and support during many observing runs at the observatory, especially Francisco Prada for observation and reduction of some bright galaxy spectra. We thank D. Calzetti for kindly making available the galaxy templates in digital form. This work was supported by the DFG (Sonderforschungsbereich 439).
\end{acknowledgements}

\appendix

\section{Notes on peculiar objects}

Here, we would like to discuss in detail some objects with peculiar properties making them a non-trivial case for the classification. Some of these are also classified as {\it strange}, while the peculiarity of others was only revealed by taking a closer look:
\begin{itemize}

 \item  The bluest object of $R<24$ in the present dataset is a compact source in the 16h-field with $B \approx 22.8$ and $R \approx 22.95$ which is classified as a star. An observation of such a blue star at this magnitude level would more plausibly be explained by a white dwarf with $M_V \sim 11$ residing at a distance of about 2.5\,kpc than by a B7V star of $M_V \sim -1$ that had to be far out at 600\,kpc distance. In fact, the object was confirmed as a white dwarf on the basis of its broad H$\beta$ absorption line during the first spectroscopic observing run.

 \item  The reddest object of $I<23$ is also a compact source in the 16h-field with $R>26$, $I \approx 22.5$ and $K^\prime \approx 18.6$. It is classified as a galaxy at $z \approx 1$ having a very red SED, but spectroscopically we identified it as an L1 star \cite{Wolf98}. Obviously, the best fitting red galaxy template matched the observed colors better than the red-most star available in the Pickles library which is in fact of type M8. So, the problem here is an insufficient library lacking L-type stars.

 \item  The most nearby extragalactic object confirmed is a galaxy in the 16h-field residing at $z=0.035$. It is classified as a galaxy with a redshift estimate of $z=0.031\pm 0.003$. The redshift is mostly constrained by an emission line showing up in the medium-band filter 522/15 (i.e. O\,{\sc III}) in combination with the absence of any continuum drop within the filter set to be caused by a 4000\,\AA-break. Upon completion of the full data set, we should also see an H$\alpha$-line in the filter 683/18. Since the blue side of the spectrum is somewhat sparsely sampled by our filters, we expected problems for an accurate redshift estimate of galaxies at $z<0.2$, which appear to be not so common for emission line galaxies. The redshift yields a distance of 210\,Mpc with $H_0=50$\,km/(s\,Mpc). Using an aperture corrected total magnitude of $B\approx 21.85$, this galaxy has a rather moderate luminosity of $M_B \approx -14.8$.

 \item  The 01h-field contains a compact source of $R=23.4$ which appears in the filter 815/20 about three magnitudes brighter than in the neighboring bands. It is classified as a rather {\it strange} galaxy with $z_{phot}=1.78$. Although, we do not have a spectrum of this object, yet, we can probably clarify its nature: In fact, the high flux in the 815/20 filter is consistent with a very strong emission line seen by the Fabry-Perot observations ranging from 814 to 824\,nm. While the filter 815/20 suggests a total equivalent width for the contained lines of $\sim 2700\pm600$\,\AA, a line fit to the FPI fluxes yields $\lambda_{cen} = 815.0$\,nm and an equivalent width of $\sim 1700\pm500$\,\AA. In addition, we see the filter 611/16 brightened by an emission line and a multi-color spectrum suggestive of a $z \approx 0.4\ldots 0.7$-galaxy. 

The only consistent picture is an interpretation of the line in the FPI as O\,{\sc III} 5007 placing the object at $z=0.628$. The line in the 611/16 filter is then O\,{\sc II} 3727, while the line O\,{\sc III} 4959 is not covered by the FPI wavelength range, but contributes to the total flux in the filter 815/20, lifting the total equivalent width of the line pair up to $\sim 2300\pm650$\,\AA. Physically, this means the object is emitting $\sim 2 \times 10^{42}$\,erg/s in its O\,{\sc III} line from a giant H\,{\sc II} region or nuclear starburst. According to Balzano (1983), about 3\% of all field galaxies with $M_B<-17.5$ in the local universe show nuclear starbursts with more than $10^{40}$\,erg/s flux in the H$\alpha$-line, but less than 0.01\% have H$\alpha$-fluxes above $10^{42}$\,erg/s. While the flux ratio of O\,{\sc III} and H$\alpha$ in her sample shows quite a spread, H$\alpha$ tends to be stronger. We therefore conclude, that we most likely see one of those rare objects here. In our present sample of emission-line galaxies found with the FPI, this object has the most luminous line flux.

 \item  Our only present $z>4$-quasar candidate was identified by spectroscopy as an emission line galaxy at $z=0.265$. The object of $R=22.9$ is classified as a {\it strange} quasar with $z_{phot} = 4.12$, mostly because of its red B--R color and blue R--I colors in combination with a strong emission line seen in the filter 628/16 which also contributes also to the R-band flux. If the strong line flux (equivalent width $\ga 1200\,\AA$) is taken into account, the true level of the R-band continuum can be recovered and matched into a normal nearby blue galaxy spectrum. A pure broad-band survey would supposedly have picked this object as a $z>4$-quasar candidate. But our method produced it as a quasar candidate, too, although CADIS provided all relevant photometry to identify it as a low-redshift emission-line object. 

 \item  The most nearby active galaxy in our present sample is a Seyfert-1 galaxy at $z=0.474$ found by chance in the 16h-field. It is the least luminous AGN we found, having $M_B=-21.7$ derived from $R=20.0$ for $H_0=50$\,km/(s\,Mpc) and $q_0=0.5$ \cite{Wolf99}. It is compact in appearance and classified as a {\it strange} starburst galaxy. The strangest part of its multi-color spectrum is a rather bright K-band magnitude of $K^\prime=16.3$ indicating an NIR excess. In fact, at lower luminosities and redshifts we have to expect quite some incompleteness in searching for Seyfert-1 galaxies. First, our libraries do not contain any NIR excess over a pure power law and second, spectra of faint Seyferts are likely to be a superposition of a host galaxy with an active nucleus requiring some composite templates.

 \item  Another Seyfert-1 galaxy confirmed at $z=1.22$ in the 9h-field having $R=21.2$ and $M_B=-22.2$ is classified as a strong starburst galaxy from the present dataset. However, when fewer filters were available it was classified and found as a quasar. This is our only case, where a significant enlargement of available color information removed a good and proven candidate from the correct class.

 \item  Only one of the quasars we identified appears extended. It resides at $z=1.57$ in the 16h-field having $R=21.5$ and $M_B=-23.4$. We believe to have resolved a luminous host galaxy, since the redshift estimates for both the galaxy and the quasar class coincide with $z_{phot} = 1.83\pm 0.03$ and $z_{phot} = 1.63\pm 0.01$, respectively. An alternative interpretation is, that we are just looking at a chance projection of the quasar with a foreground object, but our spectrum of this object is too noisy to look more closely into that.

\end{itemize}

\end{document}